# Title: Signature of a pair of Majorana zero modes in superconducting gold surface states

Sujit Manna[1,2,†], Peng Wei[3,†*], Yingming Xie[4], Kam Tuen Law[4], Patrick Lee[1]* and Jagadeesh Moodera[1,5]*

**Affiliations:**

[1]Department of Physics, Massachusetts Institute of Technology, Cambridge, MA 02139, United States

[2]Department of Physics, Indian Institute of Technology Delhi, New Delhi 110 016, India

[3]Department of Physics and Astronomy, University of California, Riverside, CA 92521, United States

[4]Department of Physics, Hong Kong University of Science and Technology, Hong Kong

[5]Francis Bitter Magnet Laboratory, and Plasma Science and Fusion Center, Massachusetts Institute of Technology, Cambridge, MA 02139, United States

[†]These authors contributed equally

[*]Correspondence to: peng.wei@ucr.edu, palee@mit.edu, moodera@mit.edu

**Abstract:**

**Under certain conditions, a fermion in a superconductor can separate in space into two parts known as Majorana zero modes, which are immune to decoherence from local noise sources and are attractive building blocks for quantum computers. Promising experimental progress has been made to demonstrate Majorana zero modes in materials with strong spin-orbit coupling proximity coupled to superconductors. Here we report signatures of Majorana zero modes in a new material platform utilizing the surface states of gold. Using scanning tunneling microscope to probe EuS islands grown on top of gold nanowires, we observe two well separated zero bias tunneling conductance peaks aligned along the direction of the applied magnetic field, as expected for a pair of Majorana zero modes. This platform has the advantage of having a robust energy scale and the possibility of realizing complex designs using lithographic methods.**

**Significance statement:** We found signatures of Majorana zero modes in superconducting gold surface states. This new platform holds promise for scalable Majorana qubits architectures.

**Main text:**

Majorana zero modes (MZMs) are fermionic states, each of which is an antiparticle of itself. MZMs are required to always appear in pairs.[1] Each MZM pair has the degrees of freedom of a single fermion, which is split non-locally in space into two MZMs. The nonlocality implies that MZMs are immune to local perturbations, and hence they have been proposed as the key ingredients of topological qubits that are protected from decoherence due to local noise sources.[1,2] MZMs are predicted to exist in certain topological superconductors (TSCs). Up to now, various proposals have been made to engineer TSCs by combining conventional materials,[3-9] and great progress has been made towards revealing the signatures of MZMs.[10-29] Here we report the observations of signatures of MZMs in the form of zero bias peaks (ZBP) in a new platform which is based on the surface state of gold. Our metal-based platform holds a number of advantages. First there is a wealth of experience in epitaxial growth of noble metals on a variety of superconductors,[30-32] as well as epitaxial growth of ferromagnetic insulators on top.[33] Second, lithographic methods to produce large arrays of increasingly complex designs are well developed. Finally the energy scales in a metal are generally high compared with those of semiconductors. While the eventual goal of producing a topological qubit based quantum computer is still daunting, our work opens a new vison of a path forward.

Our platform is based on the Shockley surface state (SS) of (111)-gold (Au) with induced superconductivity,[33] as proposed theoretically several years ago.[34] An important motivation of using the SS of Au(111) is its large Rashba spin-orbit coupling (SOC) energy scale leading to a splitting of 110 meV,[35,36] which is several orders of magnitude larger than those in semiconductor naonwires. A schematic drawing of the needed heterostructure is shown in Fig 1a. A gold film is grown on top of a superconductor (vanadium (V) in our case) and the bulk gold becomes superconducting due to the proximity effect. The bulk gold in turn induces a pairing gap on the surface state. We have shown earlier that the surface state has an energy gap of 0.38 meV, distinct from the bulk gold gap which is 0.61 meV.[33] On top of gold we grow an epitaxial film of EuS, a ferromagnetic insulator to magnetize the gold surface state via exchange coupling. We have created arrays of Au(111) wires (Fig 1b) fabricated utilizing our epitaxially grown thin film layers.[32,33] The wires are 4 nm thick, approximately 100 nm or less wide, and microns in length as seen in Fig 1b and 1c. We emphasize that our Au(111) wires are highly crystalline with sharp edges (Fig 1c). The scanning tunneling spectroscopy (STS) spectrum shows clearly that a hard superconducting gap is induced in the Au wire from the vanadium underneath (Fig 1d, Fig 3a and Fig 4d). The Au wire is homogeneously in contact with the vanadium layer sharing a high-quality interface that is made under ultra-high vacuum environment (Fig. 1a and see Methods). However, the proposal based on the SS of Au(111) has a serious limitation. The bottom of the SS band lies about 500 meV below the Fermi level.[35-37] For a 100nm wide Au wire, this means that there are approximately 100 transverse sub-bands crossing the Fermi level. Our theory simulations show that the topological state is extremely delicate under such condition, and it is unlikely that the MZMs have detectable signals (see SI Section 3). Another limitation is the need of a large parallel magnetic field to overcome the induced superconducting gap (Fig 1d, Fig 3a and Fig 4d), which is much larger compared to platforms based on semiconductor with large g-factors.

We overcome the above limitations by depositing a thin layer of EuS on top of the Au wire, which has been shown to be effective in dramatically lowering the Fermi level of SS.[33] In addition, EuS is a ferromagnetic insulator, known to induce sizeable magnetic exchange field in adjoining layers, thereby reducing the needs of a large applied magnetic field.[38-41] Fig 2d shows that two monolayers (MLs) of EuS deposited on the Au wire (Fig 2a-2c) is enough to shift the bottom of the SS band all the way to only ~ 30 meV below the Fermi level. We ascertain this by performing a series of STS scans along the surface of a Au nanowire and comparing them with similar scans on top of a two MLs EuS island. The STS peak marked by arrow is the signature of the band bottom of SS, characterized by a square root singularity of the density of states. The STS peak of the pristine Au nanowire surface is at about 420 meV below the Fermi energy, consistent with the values reported by angle-resolved photoemission spectroscopy (ARPES)[35,36] and STS[37]. We define the chemical potential $\mu$ as the energy difference between the Fermi level and the crossing point of the Rashba bands, which lies 15 meV above the band bottom. We conclude that two MLs of EuS places the chemical potential $\mu$ to be about 15 meV. This means that for a 100 nm wide wire, only about 5 transverse sub-bands cross the Fermi level and the condition for creating MZM is much more favorable (see SI Sections 4 and 5).

If a middle segment of the Au wire is completely covered by EuS, two MZMs are expected to emerge underneath the ends of the EuS in the presence of a magnetic field applied parallel to the wire. Unfortunately, STS probe of the Au SS through two monolayers of EuS can only be done with low energy resolution. While a continuous EuS layers can be grown to completely cover the surface of a Au wire when a thick layer of EuS is deposited, we find that EuS islands that are uniformly two monolayers thick are formed when the EuS coverage is low (see Fig. 2c). Although planar tunnel junctions can be used to tunnel through a thick EuS layer,[33] as a proof of concept, we focus on EuS islands and use STS for the tunneling measurements (Fig 2c). In order to obtain high resolution STS, we tunnel into the exposed Au(111) surface regions in the vicinity of a EuS island ~ 2 nm away from the island edge. Fig 1a shows a schematic of such measurements. If MZMs emerge underneath the EuS island, the STM tip can still couple to the MZMs due to their spatial decay (Fig 1a). The decay length is found to be about 8 nm (see SI section 8), so 2 nm is well within the decay length.

We found ZBPs emerging inside the pairing gap when an applied magnetic field exceeds a certain value. Fig 3 shows the tunneling spectra taken at a number of positions around an island about 30nm wide and 25 nm long sitting on an Au nanowire (Fig 2c). At zero field, the dI/dV spectra demonstrate a hard superconducting gap, which does not change regardless of the location where the STS is taken (Fig 3a and Fig 4d). The results demonstrate a uniformly induced superconductivity in Au(111), and the shape of the EuS island has no effect on the induced gap. With 4.8T magnetic field applied along the wire (called the north-south direction), a ZBP appear in positions 1, 2 and 3 near the north side of the island and near position 6 on the opposite side (Fig 3b). On the other hand, anywhere in between these positions, the tunneling spectra remain largely unchanged except for a small filling in the gap due to the applied magnetic field (Fig 3a). To visualize the evolution of the gap and the ZBP, i.e. to demonstrate the topological transition, we take the STS data with a fine scan of the magnetic field near position 6 (Fig 2c). In a 2D density plot, a transition characterized by the closing of the superconducting gap and the emergence of ZBP beyond a critical field value is observed (Fig 3c). The line scans (Fig 3d) show that the superconducting gap evolves from "U" to "V" shape before the gap is filled in and the ZBP emerges after that. As we shall see, these features closely resemble the topological transition in a multi-mode system modeled by our theory (Fig 6c).

We note that the width of the ZBP is about 0.2 meV, about a factor two larger than $\pi k_B T$ that is expected for the temperature of 0.38 K. The width is limited by the instrumental resolution due to the bias modulation voltage needed to achieve a signal with a sufficiently low noise (See Method). Therefore, it is possible that the ZBP are due to a pair of MZMs which are hybridized to give a pair of split peaks that are not resolved. In fact, such splitting is to be expected for an island that is so small.

We next show results on a more rectangular shape island (60nm×45nm) with an edge which overlaps the edge of the Au(111) wire as shown in Fig 4b. The EuS island has excellent crystalline quality as can be seen from the atomically resolved scanning tunneling microscope (STM) image (Fig 4c). We apply a magnetic field parallel to the wire and take STS measurements in the exposed Au(111) regions around the island. Again, we find uniform and fully gapped spectra at zero field (Fig 4d). In 4.8T field, two ZBPs are observed on the opposite sides of the EuS island at the positions where the island intersects the edge of the gold wire (Fig 4b and 4e). At all other positions (labeled 2-7), the superconducting gap spectra remain unchanged except for a slight filling of the gap. Apparently the wavefunction corresponding to the ZBPs are strongly localized along the intersected edge of the EuS island and the gold nanowire, rather than spread out along the upper and lower edges of the EuS island. This observation is also found in our simulations discussed below, as shown in Fig 5a. The localization of the MZM is sensitive to the local environment such as steps in the chemical potentials at the edge of the gold wire, thus accounting for the strong localization in this case.

We next study the evolution of ZBPs at positions 1 and 8 as a function of the applied magnetic field (H). At both positions the gap fills in when H is around 3.5T (Fig 4f and 4g), which agrees well with the onset of ZBPs in the other island (Fig 3c). The ZBP appears at comparable threshold magnetic fields for position 1 (at ~ 3.5T) and for position 8 (at ~ 4T) (Fig 4f and 4g). The slightly different threshold magnetic fields can be a result of the broad transition region as seen in Fig 3c. For Fig 3c, the transition takes place at around 3.5T but has a blurred region with a span of the magnetic field of more than 0.5T. We note that a sharp threshold field and the emergence of MZMs simultaneously at the same threshold field are not expected under our experimental conditions. There are two reasons. The first is that unlike the quantum wire case, we are in a multi-mode situation. As explained below and supported by simulations shown in Fig 6, the gap closing and the emergence of the MZM take place rather gradually due to a multitude of low-lying states near the threshold. The second reason has to do with quasi-particle poisoning due to excitations above the topological gap (see Fig. 5b) which is defined as the energy of the lowest extended quasi-particles state. This gap closes at the topological transition and re-opens at higher fields, so that just above the transition we expect a large poisoning rate $\Gamma_p$. The weight of the ZBP is reduced by the factor $\Gamma/(\Gamma+\Gamma_p)$ where $\Gamma$ is the tunneling rate.[42] Since $\Gamma$ is very small for STM tunneling, a large reduction of the ZBP weight is expected near the topological transition at around 3.5T. The weight gradually increases as the topological gap reopens and $\Gamma_p$ is reduced. Since $\Gamma$ and $\Gamma_p$ depends on the local details near the location of the MZM, the visibility and the onset magnetic field of each MZM partner is expected to have some variation, consistent with Fig 4f and g.

To support our interpretation, we have carried out numerical simulations on realistic models of the Au(111) wire partially covered by EuS. We perform the calculation on a square lattice with hopping and Rashba energies chosen to match the band structure of the gold surface states. We allow the chemical potential to be reduced and the effective Zeeman energy $V_x$ to be enhanced under the EuS island. The coupling between the surface state and the bulk gold states is taken into account with a self-energy correction (See SI Section 1). Indeed we found that when the chemical potential under EuS is less than 30 meV, the MZMs exists over a large fraction of the chemical potential range with a relatively large gap to other sub-gap excitations, so that the chance of finding robust MZMs is quite high (see SI Fig S2b,c). The spectrum and the distribution of the wave-function are shown in Fig 5 for an island that is 60nm×60nm which overlaps the edge of the gold wire. In Fig 5a we see that the MZM wavefunctions are strongly localized at the corners of the intersected edge, in excellent agreement with what is seen experimentally in Fig 4. In contrast, for a rectangular shaped island that is in the middle of a gold wire and removed from the edge, the MZM wavefunctions are spread out along the north and south edges (See Fig. S3). It is apparent that the MZM wavefunction can be trapped by spatial inhomogeneity such as the proximity to a step in the chemical potential near the edge of the gold wire. A second point we find from the calculation is that the decay length of the MZM perpendicular to the edge towards the middle of the wire is very short. This explains the surprising fact that ZBP were observed for islands as small as 30nm in diameter. The coherence length is short for two reasons. First, the Fermi velocity is very small because we are considering the last filled sub-band which has a very small Fermi energy. Secondly, this situation is similar to what was observed in the Fe atomic chain,[16] where the wavefunction leaks into the substrate, giving rise to a reduction of the spectral weight and the velocity, thereby reducing the effective coherence length.[43] In fact, this leakage is needed to produce a surface superconducting gap close to the bulk superconducting gap $\Delta_B$.[34]

A key ingredient in the simulation is the assumption of an exchange field under the EuS in addition to the applied magnetic field. Without this, a magnetic field large enough to drive the system to a topological regime also closes the superconducting gap outside and the MZMs will be destroyed. In EuS the magnetic moment normally lies in-plane, but on certain surfaces it is known to develop a canted magnetization with large out of plane component due to spin-orbit coupling.[44,45] In this case the increasing applied parallel magnetic field enhances the Zeeman field along the nanowire due to the canting of the magnetic moment. However, the precise orientation of the magnetic moment in our case is not known.

We note that much of the literature deals with the case with a single transverse mode which is relevant to the semiconductor nanowires. Here we highlight the difference with the multi-mode situation that is relevant to our system. To make the physics clearer, we focus on a longer island that is 60nm × 300nm where the energy splitting of the MZM is small and well separated from the higher energy states. Fig 6a shows the lowest 25 eigenvalues as a function of the effective Zeeman energy $V_x$. We see that the gap defined by the lowest excited state closes at $V_x = 1.1\ \Delta_B$ and reopens, leaving behind a pair of split MZMs. The topological gap that re-opens is about $0.2\ \Delta_B$. Note that a large number of states come down in energy as the gap closes and a large number lies above the topological gap that re-opens. Consequently, the contribution of an individual state to the tunneling conductance is very small. In the single mode case the number of states that come down as the gap closes is much fewer, depending on the sample length, and can even be a single state for short wires (see Fig S2f,g). As a result, the gap closing as the topological state is approached can be seen both in experiment and simulation in short nanowires.[23] In contrast, in our case the lowest state that leads to gap closing as the threshold is approached contributes only a small amount to the tunneling conductance (Fig 6a). Upon thermal smearing, we expect a subtle filling-in of the gap which becomes V shaped, as clearly seen in the right panel of Fig 6c where we show the calculated tunneling spectrum at 700 mK to simulate the additional instrumental broadening. The V shaped spectra in the transition region also agree well with our experiment (Fig 3d). With further increase of $V_x$ beyond the threshold, the gap fills in more and eventually the MZM emerges as the topological gap opens. We emphasize that this behavior is intrinsic and not the result of disorder. We have also examined the tunneling conductance of the gold surface state at a point halfway between where the ZBP appears, and just outside of the EuS island. Only the extended states contribute to these spectra. At 350mK, we can see a slight and gradual filling in of the gap as the Zeeman field reaches and exceed the threshold field of $V_x = 1.1\ \Delta_B$ (see Fig. 5d), in qualitative agreement with what is shown in Fig 4d.

Next, we address the question of whether the ZBP we observed may be due to conventional Andreev bound states (ABS) that happens to stick to zero energy. Two sources of such ABS have been discussed depending on whether disorder is present. In the absence of disorder, it has been pointed out that if the end of the wire is subject to a smooth potential, even in the non-topological region a pair of MZMs are often found near the end that are weakly hybridized with each other.[46-49] One of these could couple strongly to the outside lead and the tunneling conductance resembles that of a true MZM. The reason for the lack of hybridization is that the two states may have different spins or momenta and are therefore almost orthogonal.[48,49] Such smooth potentials are generic in the semiconductor wire set-up because the tunnel barrier is created by gating. In our case, the wire and the island terminate abruptly on an atomic scale and the tunnel barrier is via an STM tip. Therefore, we do not expect this mechanism to be applicable in our system, and indeed we do not see any sign of such false signals in our simulations.

A second source of false signal is that disorder potential may accidentally produce a localized Andreev bound state that happens to have energy near zero. However, such a mechanism usually requires fine-tuning of the parameters. Furthermore, it is equally likely to find such a localized state anywhere along the boundary of the island. This is in contrast to our findings because our ZBPs are aligned with the applied magnetic field and we see no sign of in-gap peaks in zero or finite field anywhere else. Finally, there is no reason why these Andreev bound states will appear at two areas on opposite sides of the same island at a similar magnetic field. In this connection we mention that the appearance of ZBP's in the EuS islands is not a rare occurrence. We have studied about 60 islands under varying experimental conditions which are sometimes not ideal in terms of STM stability. On 20 of these islands we found ZBPs which are always located either near the north or south shore with respect to the magnetic field. We also note that the spectra we see are either fully gapped or exhibit ZBPs and we never find split peaks that may be associated with Andreev bound states. Out of the 20 we saw a pair of ZBP's at opposite ends on 4 islands, even though in many cases we simply did not explore all the way around the island. In any event, since we do not control the chemical potential, we expect to find MZM less than 50% of the time based on our simulations, consistent with our finding.

Our simulation also found that an EuS island in the middle of a wider Au wire and away from the edges can also support MZMs, even though the wavefunction is more spread out along the edge of the island (Fig S3) and thus will be harder to detect with STM. We believe this explains the observation of ZBP localized on the edges of the island shown in Fig 2 and Fig 3. Importantly, this observation opens the door to the possibility of depositing EuS wires on a large area of Au which will allow even greater freedom in designing more complicated structures towards MZMs based topological qubits. Obviously, this is something we plan to explore in the future.

The gold surface state platform has a high intrinsic Rashba scale and there is room for further improvement. In the future by going to planar tunneling, an order of magnitude reduction in temperature and energy resolution should be possible. A larger magnetic field can be applied which allows us to utilize superconductors with large gaps, resulting in larger topological gaps. Any coherent manipulation of MZMs requires the topological gap to be much larger than temperature. The gold surface state platform holds a lot of promise towards this goal,

The Au/EuS system is amenable to flexible nano-fabrication methods. For example, a single topological qubit can be constructed out of two or more wires that are cross-linked, forming "tetron" or "hexons" that have been proposed to be building blocks of quantum computing circuits.[50,51] An array of such objects can be constructed without too much difficulty (see SI Section 7). In particular, instead of braiding by physically moving MZMs, a measurement based topological qubit design using a purely superconducting circuit is well suited with our Au/EuS system.[52]

In conclusion, we have found signatures of MZMs pairs on superconducting Au surface as theory predicted. The Au/EuS platform has the advantage of having a robust energy scale and the potential for fabricating a large-scale network. This platform offers a plausible path, though still an extremely challenging one, towards topological quantum computing.

**Methods**

The heterostructures of thin film V(20 nm)/Au(4nm) are grown in a custom-built molecular beam epitaxy (MBE) system with a base vacuum ~ $3.7 \times 10^{-10}$ torr. The growth and characterizations follow the similar procedures as we reported before.[32,33] Au nanowires, or nanoribbons, with a thickness of 4 nm and a wire width around 100 nm (or less) are fabricated after the growth of Au/V heterostructures. The nanowires are sculptured out of the wafer-scale Au thin film by following a standard e-beam lithography and Ar ion milling process. During the nano fabrications, the surface of the superconductor vanadium is protected in-situ using a custom-built high vacuum system (base vacuum $10^{-8}$ torr) that combines both the ion milling and thin film evaporation sources, thereby guaranteeing the high quality of the superconductor V. Any shape of nanowire, or a network of nanowires, can be designed by standard e-beam lithography software and fabricated following such an approach (see SI Section 7). The surfaces of Au nanowires fabricated in this way are ultra-clean as seen by the atomic resolution images of the STM (Fig. 1). A post growth of EuS magnetic insulator is carried out in a high vacuum system with a base vacuum ~ $10^{-8}$ torr. The EuS growth, from islands to a continuous layer, on Au(111) surface is controllable by varying the EuS thickness as described before.[33] EuS islands are chosen here to facilitate the STM measurements, because the EuS thick films are too insulating. The STM and STS experiments are performed in a UHV custom assembled STM with RHK PanScan head integrated in a Janis 300mK He3 cryostat with a 5T vector magnet. The STM is equipped with a rotating flange that allows a precise rotation of the scanning head with respect to the magnet of the cryostat, allowing us to align the applied magnetic field with the Au nanowire. Tunneling spectroscopy measurements were taken using a standard lock-in technique at 937 Hz frequency and bias modulation voltages of $V_{mod}$ = 80-120 μV (RMS value). The STM tips used were home-made, either chemically etched bulk W tips or commercial bulk PtIr tips.

**Acknowledgements:** We thank Yunbo Ou, Mirko Rossi, Yota Takamura, and Juan Pedro Cascales Sandoval for technical help in conducting the experiment. P.W., S.M., P.A.L., and J.S.M. acknowledge support from John Templeton Foundation grants 39944 and 60148. P.W., S.M., and J.S.M. acknowledge Office of Naval Research grants N00014-13-1-0301 and N00014-16-1-2657 and National Science Foundation grants DMR-1207469 and DMR-1700137. P.W. and J.S.M. acknowledge National Science Foundation QLCI-CG grant 1937155. P.A.L. acknowledges

Department of Energy (DOE) grant DE-FG02-03ER46076. K.T.L. was supported by the Croucher Foundation, the Dr. Tai-chin Lo Foundation, and Hong Kong Research Grants Council grant C6026-16W.

**Author contributions:** P.W., P.A.L., and J.S.M. designed research; S.M., P.W., Y.X., and K.T.L. performed research; S.M., P.W., P.A.L., and J.S.M. analyzed data; Y.X., K.T.L., and P.A.L. performed theory modeling; and P.W., P.A.L., and J.S.M. wrote the paper with contributions from all authors.

# Figures

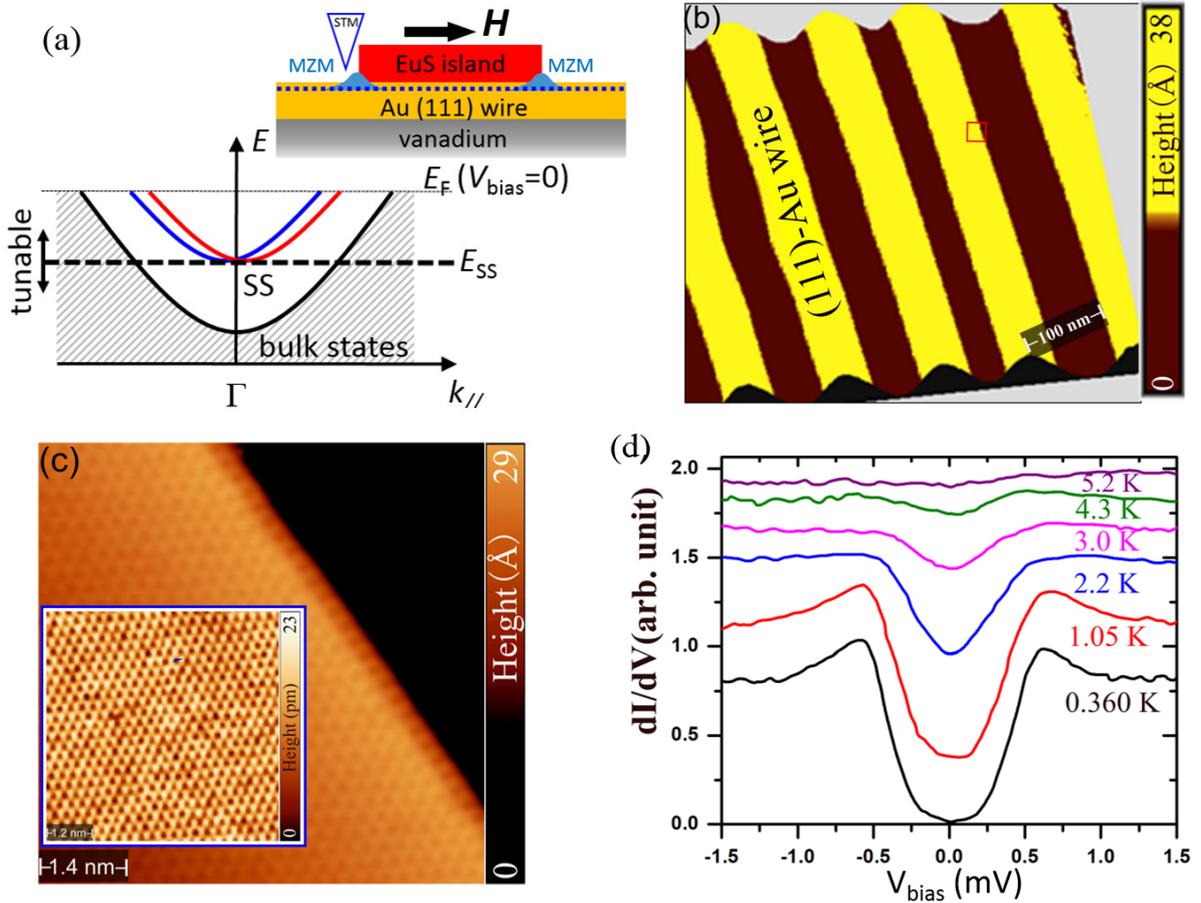

**Figure 1.** (a) Schematic of the experimental setup: Au(111) thin film nanowire proximity coupled to a conventional superconductor vanadium, while the EuS is grown epitaxially on top of the Au nanowires. Dotted line represents the location of the surface state (SS). An external field is applied parallel to the wire in order to drive the system into a topological superconducting state. A scanning tunnelling microscopy (STM) with a normal tip is used to probe the part of the MZM that leaks out from underneath the EuS island. Also shown is the schematic surface state Rashba split band structure which is isolated from the projected bulk bands. The position of the bottom of the surface band ($E_{SS}$) can be tuned by varying the thickness of the EuS coverage. (b) Large scale (650 × 650 nm$^2$) STM with constant current ($V_{sample}$= 1.2 V, $I_{set}$ = 55 pA, T = 2 K) topographical scans of the nanowires network that is prepared using nanofabrication techniques. (c) A zoomed-in (7nm × 7nm) (shown by square box in (b)) topography of Au nanowire which showed sharp interface with the underlining vanadium film ($V_{sample}$ = 0.36 V, $I_{set}$ = 230 pA, T = 2 K). Inset shows the atomically resolved STM topography image ($V_{sample}$ = -150mV, $I_{set}$ = 0.6 nA, T = 2 K) of the Au nanowire top surface, which shows the hexagonal atomic lattice of Au(111) surface (d) Temperature dependent dI/dV spectra measured on atomically resolved Au nanowire surface. The spectra are spatially averaged over a 1.5 nm × 1.5 nm area located at the bottom right corner of Fig. 1c inset. The spectra are vertically shifted and normalized by the data measured above $T_c$ (~ 5 K).

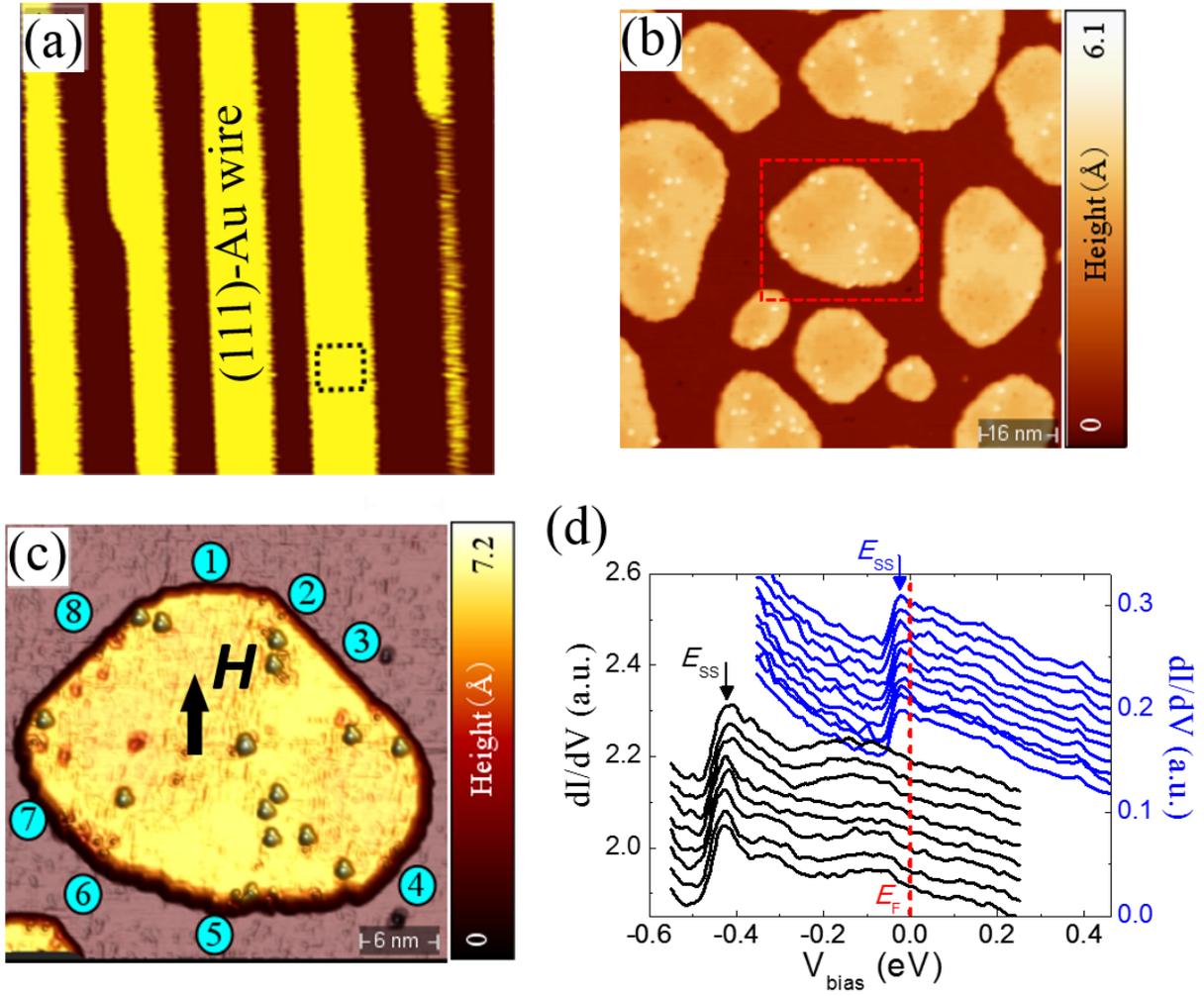

**Figure 2.** (a) Large scale STM topography image of the Au(111) nanowire array (similar to that in Fig. 1b) with two monolayers of EuS deposited over it ($V_{sample}$= 1.0V, $I_{set}$ = 80pA, T =0.38K). The size of the marked square is the same as the size of (b). (b) The zoomed-in STM topography image (80 x 80 nm$^2$, $V_{sample}$= 0.8 V, $I_{set}$ = 210 pA, T = 0.38K) of the area denoted by a square mark in (a). The EuS islands are clearly visible. (c) The further zoomed-in STM topography image (35 x 35 nm$^2$, $V_{sample}$= 0.5 V, $I_{set}$ = 0.5 pA, T = 0.38K) of a specific EuS island shown in (b). (d) The STS (dI/dV vs $V_{bias}$) spectra over large bias voltages on both bare Au(111) surface (black) and EuS island (blue). A clear shift of $E_{SS}$ (bottom of the surface band) towards $E_F$ is seen, which shows that the Fermi level of the gold surface state underneath two monolayer EuS is ~ 30 meV relative to $E_{SS}$. Here we emphasize the homogeneity of the sample, as demonstrated by the weak position dependence of the spectra as shown by the multiple dI/dV scans along a line. The curves are shifted vertically for clarity.

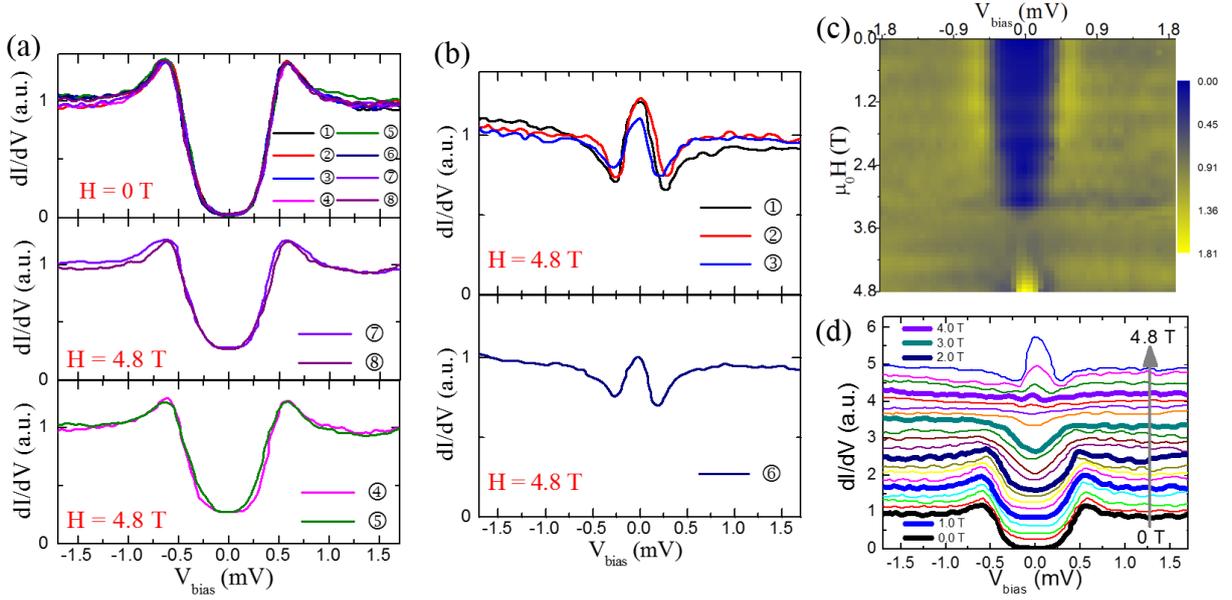

**Figure 3.** (a) The tunneling dI/dV spectra at different positions (positions 1 through 8) surrounding the EuS island acquired at T = 0.38 K as shown in Fig 2c. Top panel shows that when the applied magnetic field is zero ($H$ = 0T), the dI/dV spectra at all positions are fully gapped. The lower panels show that in a magnetic field ($H$ = 4.8T) the dI/dV spectra at position 4, 5, 7, 8 are partly filled in, but the coherence peak at the pairing gap remains intact. The direction of the magnetic field is shown in Fig 2c. (b) Sharp ZBPs emerge at positions 1, 2, 3 and 6 consistent with the expected appearance of a pair of MZM on opposite ends of an EuS island as defined by the magnetic field. (c) The 2D density plot of the dI/dV data showing the topological transition of the MZM. The data are taken at a location near position-6 of the island shown in Fig 2c. When the applied field increases, the superconducting gap is filled in, then it crosses over to a ZBP at sufficiently large magnetic field. (d) The detailed line scans of (c) are shown for various applied magnetic field changing in small steps from 0 T to 4.8 T. The dI/dV spectra are vertically shifted for clarity. Before shifting, each spectrum is normalized to 1 at the largest $V_{bias}$. Starting from the bottom spectrum (field is 0 T), each vertically shifted spectrum is for a magnetic field increased in steps of 0.25 T, except for the last spectrum (the topmost one) which has an applied field of 4.8 T. In this detailed line scan plot, one can clearly see that the spectrum evolves from "U" to "V" shaped before completely closing, and a ZBP emerges above the filled-in gap. These data resemble our theoretical results shown in Fig. 6c with T = 700 mK.

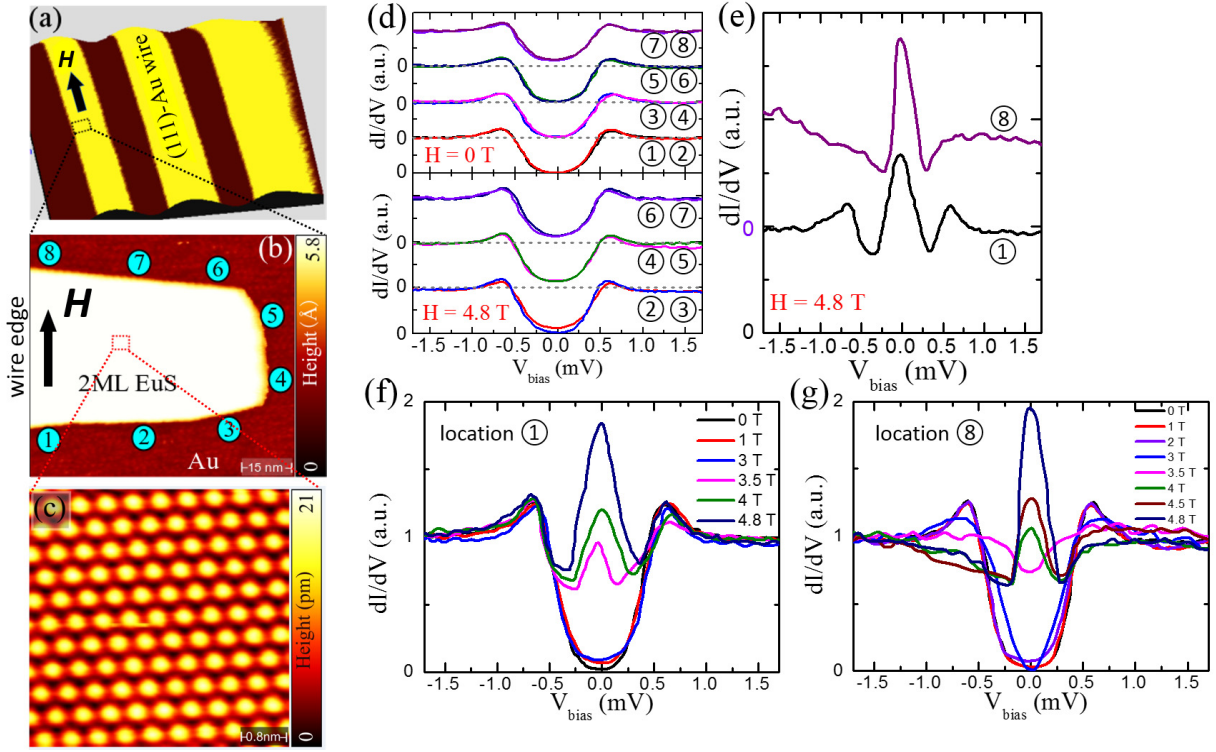

**Figure 4.** (a) The STM topography image of another Au(111) nanowire array (similar to that in Fig. 2a) with two monolayers of EuS deposited over it ($V_{sample}$= 1.6 V, $I_{set}$ = 46 pA, T = 0.38K). The applied magnetic field is aligned with the nanowire to the best accuracy of our STM setup. (b) The zoomed-in STM topography image (75 x 75 nm$^2$, $V_{sample}$= 1 V, $I_{set}$ = 0.5 pA) of a relatively large EuS island sitting at the edge of the Au nanowire in (a). The island is approximately 40nm long along the wire and 60nm wide. (c) The atomically resolved (4 x 4 nm$^2$, $V_{sample}$= 120 mV, $I_{set}$ = 600pA) EuS surface in the marked region as noted in (b). (d) The comparison of the dI/dV tunneling spectra under both $H$ = 0 T and $H$ = 4.8 T. Dashed lines mark the zero conductance of each shifted spectrum. The spectra at all positions are gapped when $H$ = 0 T. A slight filling in of the gap is seen at positions 2-7 in 4.8T field. (e) Sharp ZBP emerges for $H$ = 4.8 T at positions 1 and 8. Curve 8 is shifted vertically by two tick marks. The zero conductance for each curve is noted with the same color as the curve. (f) and (g) show the evolution of the dI/dV spectra at position 1 and 8 as a function of the strength of the applied field. At 3.5T the gap is largely filled in at positions 1 and 8 simultaneously. The ZBP is visible at 3.5T at position 1 and at 4T at position 8. At 4T field and above, the ZBP's at positions 1 and 8 show comparable peak heights. All the dI/dV spectra in (d)-(g) are normalized to the normal state conductance.

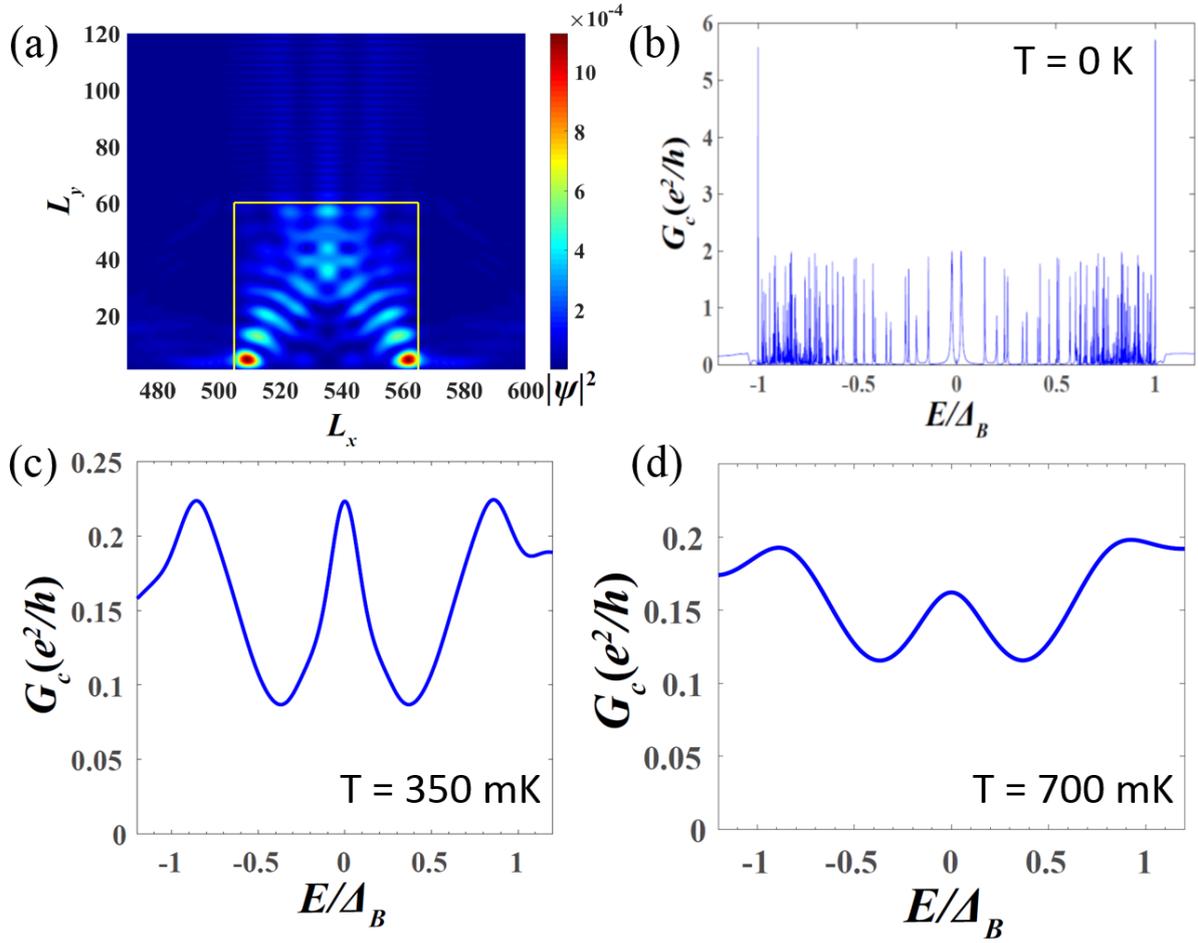

**Figure 5** Simulation for a 60nm×60nm island of EuS (yellow box in (a)) deposited on a 120nm wide Au wire. The island overlaps the lower edge. Under the island the chemical potential μ=25meV and the effective Zeeman energy is $V_x = 2\Delta_B$. In (a) the absolute value of the MZM wavefunction is found to be strongly localized on the edge, in agreement with the data shown in Fig 4. The tunneling spectra at T=0 (b) and 350mK (c) and 700mK (d) are taken at a point near the edge just outside the EuS, showing a ZBP due to a MZM. The 700mK curve is chosen to mimic the additional instrumental broadening and can be compared with the experimental curves shown in Fig 4e. The split peak near zero bias in (b) is due to the overlap of the MZM localized at opposite ends of the islands. In addition, there are in-gap states which typically start at energy of 0.1 $\Delta_B$. Thermal smearing picks out these states as sidebands and background, as shown in (c) and (d).

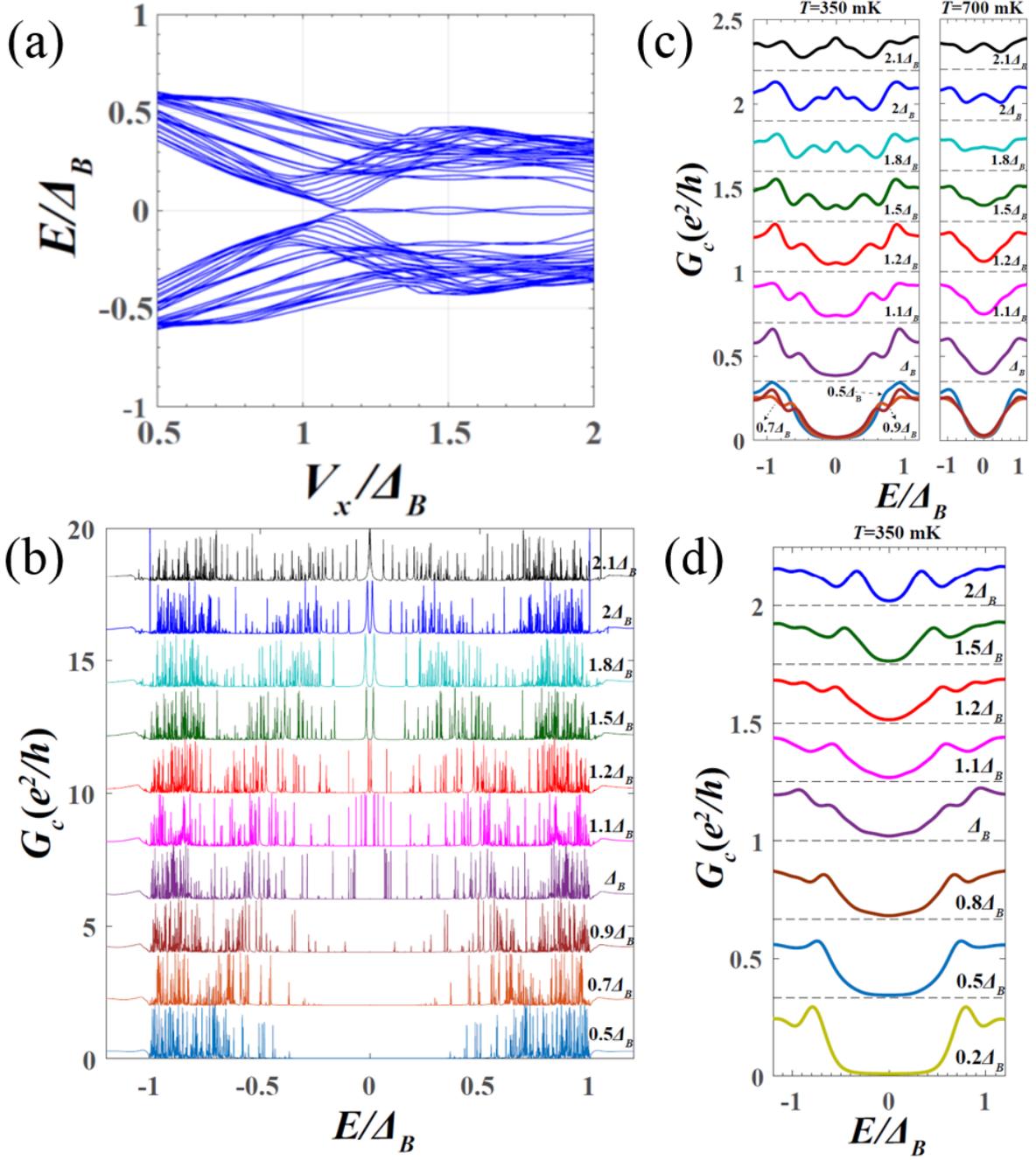

**Figure 6** Simulations for a 60nm×300nm island of EuS that overlaps the edge of a 120 nm wide Au wire. (a) The lowest 25 eigenvalues plotted vs the effective Zeeman energy $V_x$. The gap closes near $V_x = 1.1\Delta_B$ and re-opens to a topological gap of about $0.2\Delta_B$. In (b) and (c) the tunneling spectra taken near the edge just outside the EuS are shown at T=0, 350mK and 700mK. By looking at the T=0 spectra in (b) we can follow the gap closing as $V_x$ approaches the threshold. The lowest excited state carries a very small weight and is hardly visible. Above the threshold the MZM

appears as a double peak split by the overlap in a finite size island. At the same time, the topological gap develops, increasing to $0.2\Delta_B$ consistent with (a). However, a large number of states appear above the topological gap. Further examination shows that some of these states are extended, while others are localized near the ends of the wires, with a wavefunction very similar to that of the MZM. We interpret these states as "Majorana descendants", a feature of the multi-band system as detailed in SI. Upon thermal averaging, we see in (c) that at 350mK the gap closing is indicated by a rather subtle filling in of the gap and the MZM is gradually becoming visible above the threshold effective Zeeman field. The combination of extended states and Majorana descendants give rise to the sidebands of the MZM peak. In the right panel, the 700mK curves simulate the effect of additional instrumental broadening and can be compared with the measurements shown in Fig 4 f and g. (d) The tunneling spectra taken at a point at the middle of the wire just outside the EuS boundary. Only the extended states contribute. Note that the gap closing and re-opening is hardly observable and is consistent with what is seen in Fig 4d lower panel.

# Supplementary Materials for "Signature of a pair of Majorana zero modes in superconducting gold surface states"

## 1 Model

Strong spin-orbital coupling is important in realizing topological superconductivity and Majorana fermions. Potter and Lee[1] proposed a feasible scheme to realize Majorana fermions using the surface states of heavy metal thin film, which often has a large Rashba splitting. The effective Hamiltonian describing the surface states is

$$H(\mathbf{k}) = \left[\xi(\mathbf{k}) + \alpha_R (\mathbf{k} \times \boldsymbol{\sigma}) \cdot \hat{z}\right]\tau_z + \mathbf{V} \cdot \boldsymbol{\sigma} + \Delta\tau_x , \quad (S1)$$

where the basis is $(c_{\mathbf{k}\uparrow}, c_{\mathbf{k}\downarrow}, c^\dagger_{-\mathbf{k}\downarrow}, -c^\dagger_{-\mathbf{k}\uparrow})^T$, $\tau, \sigma$ operate on particle-hole and spin space respectively, $\xi(\mathbf{k}) = \frac{k^2}{2m} - \mu$ is the kinetic term of gold surface states, $\alpha_R$ is the Rashba velocity characterizing the strength of spin-orbital coupling, $\mathbf{V}$ is the effective Zeeman field, and $\Delta$ is the surface states pairing gap which is induced by the bulk-surface states mixing. The corresponding tight binding Hamiltonian can be written as:

$$H = H_0 + H_\Delta , \quad (S2)$$

$$H_0 = \sum_{\mathbf{i},\mathbf{d}} c^\dagger_{\mathbf{i},\alpha}\left(-t\delta_{\alpha\beta} + \frac{i}{2}\alpha_R(\boldsymbol{\sigma}_{\alpha\beta} \times \mathbf{d}) \cdot \hat{z}\right)c_{\mathbf{i}+\mathbf{d},\beta} + \sum_{\mathbf{i}} c^\dagger_{\mathbf{i}\alpha}\left((4t - \mu_\mathbf{i})\delta_{\alpha\beta} + \mathbf{V}_\mathbf{i} \cdot \boldsymbol{\sigma}_{\alpha\beta}\right)c_{\mathbf{i},\beta} \quad (S3)$$

$$H_\Delta = \sum_{\mathbf{i}} \Delta c^\dagger_{\mathbf{i}\uparrow} c^\dagger_{\mathbf{i}\downarrow} + h.c.. \quad (S4)$$

Here, $\mathbf{i}$ labels the position of sites, $\alpha, \beta$ label the spin, $\mathbf{d}$ is the vector connecting the nearest neighbor sites. For the case when the gold surface is partially covered by EuS, $\mu$, $\mathbf{V}$ can be spatially non-uniform. In this Rashba wire model, the Hamiltonian $H$ only has particle-hole symmetry and it belongs to topological D class. In the quasi-one-dimensional limit, namely, a strip of gold thin film, the system can be topologically nontrivial when an odd number of sub-bands are occupied. In this work, we focus on the gold surface states which have a Rashba splitting of 110 meV at the Fermi level.[2] We set $t = 16$ eV·Å$^2/a^2$, $\alpha_R = 0.4$ eV·Å$/a$ in the calculation, which have been chosen to recover the realistic continuum band dispersion[2]. The lattice constant $a$ is chosen as 1 nm for convenience. We can further estimate the important scales with these parameters, such as the spin-orbital coupling energy $E_{so} \sim 2m\alpha_R^2/\hbar^2 = \alpha_R^2/t = 10$ meV, the spin-orbital scattering length is $l_{so} \sim \hbar^2/2m\alpha_R = at/\alpha_R \sim 40$ nm.

The above model includes a pairing gap $\Delta$ in the surface states. Since we assume no direct hybridization between the surface and bulk bands, as is the case for the (111) Au surface state, the pairing gap is induced microscopically from the bulk gap $\Delta_B$ via the bulk-surface mixing due to impurity scattering, or virtual scattering via phonon or Coulomb interaction, which will introduce a self-energy term into the surface-state Green's function.[1] We will see that this self-energy term will renormalize the Fermi velocity. One of the important consequences is the localization length $\xi = v_F/\Delta$ of Majorana zero modes (MZM) can be reduced[3]. After incorporating the self-energy term, the surface-state Green's function is[1]

$$G_s(i\omega_n) = \frac{Z(i\omega_n)}{i\omega_n - \mathbf{V}\cdot\boldsymbol{\sigma} - Z(i\omega_n)(\xi(\mathbf{k}) + \alpha_R(\mathbf{k}\times\boldsymbol{\sigma})\cdot\hat{z})\tau_z - (1-Z(i\omega_n))\Delta_B\tau_x}, \quad (S5)$$

where $\Delta_B$ is the bulk pairing gap and the quasiparticle weight

$$Z(i\omega_n) = \left(1 + \frac{\Gamma}{\sqrt{\Delta_B^2 + \omega_n^2}}\right)^{-1}. \quad (S6)$$

$\Gamma$ is the strength of bulk-surface states mixing which we assume for simplicity to be due to impurity scattering, and $\omega_n$ is the Matsubara frequency ($i\omega_n = \omega + i\delta$). The simulations reported in this paper are based on Eq. S5 and S6 and we will give a short derivation of these equations in section 6. Note that we have absorbed the renormalization of the exchange field into the definition of $V$ and its x component $V_x$ under the EuS will be used as a parameter in this paper. This is explained more fully at the end of section S6. There are two important physical consequences for $\Gamma$ or the self-energy term. First, it generates the surface-state pairing gap that is given by the smallest pole in equation S5.[1] In the weak coupling region, $\Gamma \ll \Delta_B$, the surface gap is roughly $\Gamma$. In the strong coupling limit, $\Gamma \gg \Delta_B$, the asymptotic behavior of the surface gap is $(1 - \frac{\Delta_B^2}{\Gamma^2})\Delta_B$, which is comparable to the bulk pairing gap.[1] Second, it renormalizes the Fermi velocity $v_F \to Zv_F$, which reduces the localization length $\xi$.[3] This effect has been used to understand the short localization length of MZM observed in previous magnetic atom chain experiments. We indeed find that this effect plays an important role as well in localizing Majorana pairs in the partially covered island. In that case, the real space signature of MZM is revealed by solving the Rashba wire model, including the self-energy at $\omega \to 0$. We find that the choice $\Gamma = 3\Delta_B$ accounts well for both the surface gap and the short coherence length.

## 2 Method of STM tunneling spectra calculation

The tunneling conductance between STM lead and gold surface is calculated using the scattering matrix approach[4,5]:

$$G_c(E,T) = \frac{e^2}{h}\int dE'\left(-\frac{\partial f(E',T)}{\partial E'}\right)\text{Tr}[I - r_{ee}^\dagger(E')r_{ee}(E') + r_{he}^\dagger(E')r_{he}(E')], \quad (S7)$$

where $f(E,T)$ is the Fermi distribution function. An electron coming from the STM lead will be scattered when it tunnels into the gold surface. The reflection matrix in Eq. S7 can be understood as for an incoming electron, there is a $|r_{ee}|^2$ chance to be scattered back as an electron and a $|r_{he}|^2$ chance to be scattered back as a hole. Near zero bias, the Andreev reflection process will be resonant for Majorana fermions, where $|r_{he}|^2$ is 1 and the tunneling conductance $G_c$ is quantized as $2e^2/h$ at zero temperature. Technically, the reflection matrix can be calculated using the recursive Green's function method[6–8]

$$r_{ab}(E) = -I\delta_{ab} + i\tilde{\Gamma}_a^{1/2}(E) G_{nn}^R(E) \tilde{\Gamma}_b^{1/2}(E). \tag{S8}$$

Here, $a, b \in \{e, h\}$, the broadening function $\tilde{\Gamma}(E) = i(\Sigma(E) - \Sigma^\dagger(E))$, where $\Sigma(E)$ is the self-energy of semi-infinite STM lead. $G^R(E) = 1/(E + i\eta - H)$ is the retarded Green's function. $n$ labels the position where STM tunnels. This method can be easily extended to the case where the self-energy is included in the surface states.

## 3 MZM in a long gold wire

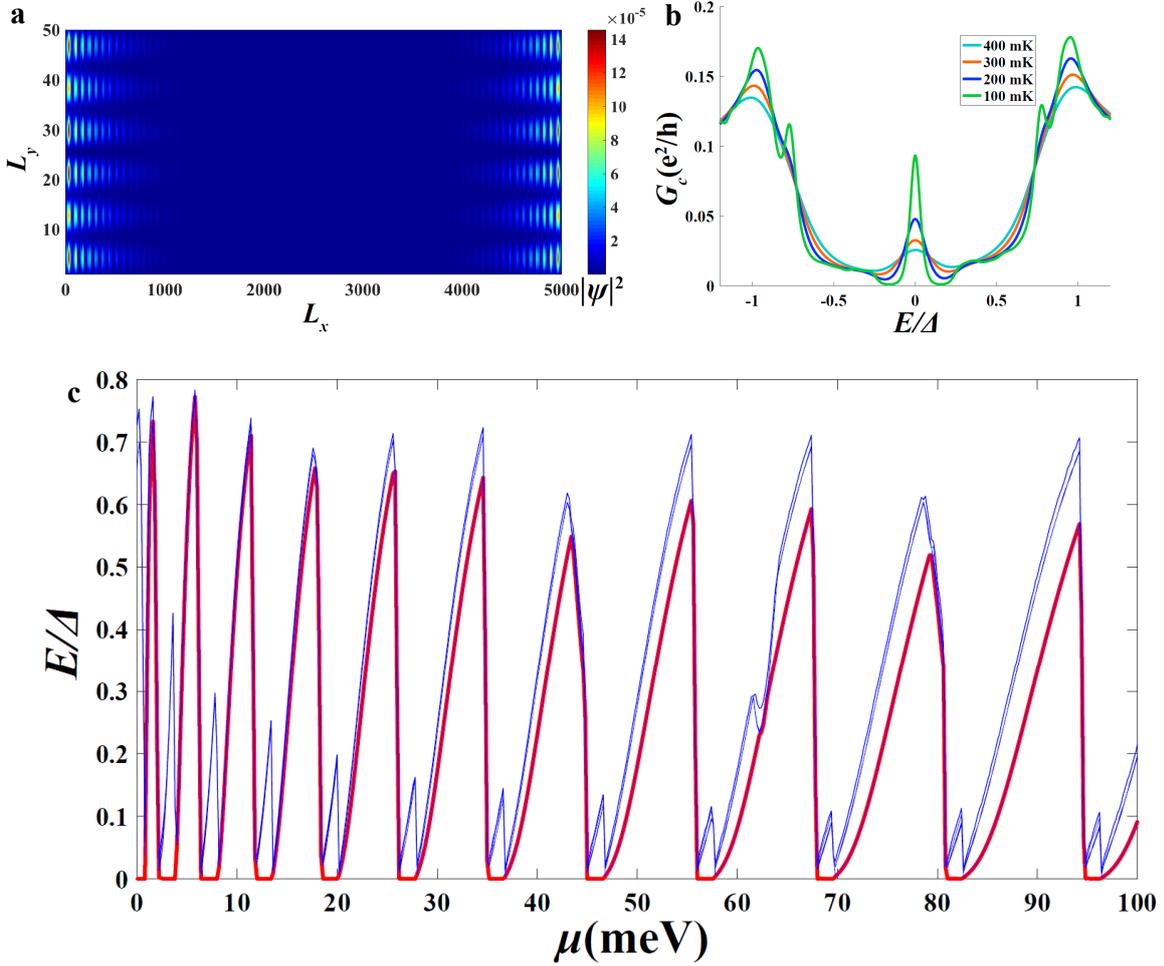

**Figure S1 | MZM and tunneling spectra in a bare gold nanowire.** The calculation is performed with tight binding model S2. The magnetic field is applied along the wire. The Zeeman energy $V_x = 2\Delta$. **(a)** The wavefunction of MZM. $L_x = N_x a, L_y = N_y a, a = 1$ nm. Here $N_x = 50$ and and $\mu$ is set to be 20 meV. The localization length looks quite long, about 100 nm, but it can be renormalized to be much shorter when we take the self-energy into consideration later. **(b)** Tunneling spectra calculated by attaching a STM lead at the end of the gold wire in **(a)**. The zero-bias peak (ZBP) is due to MZM. The height of ZBP depends on both the tunneling strength and temperature, but the relative height with respect to the background is independent of the tunneling strength because both are proportional to the square of the tunneling matrix element when the tunneling rate is much less than $k_B T$. **(c)** A plot of quasiparticle excitation energy $E$ versus $\mu$. The MZM appears (zero energy part of red line) when $\mu$ cuts through an odd number

of sub-bands. We note that the maximum topological gap $E_G$ within the topological region, which is essential for the topological protection of MZM, decreases as $\mu$ increases. As a result, a small chemical potential is essential for observing a well-protected MZM.

In this section we show the results of the simulation of the gold wire based on the tight binding model given in Eq. (S2) In Fig S1, we show the results for a gold wire with width 50 nm. At $\mu$ =20 meV and $V_x$=2 $\Delta_B$ , we find MZM's and the tunneling spectra at various temperatures are shown in Fig S1b. The MZM wavefunction is shown in Fig S1a. Fig S1c shows the five lowest eigenvalues as a function of $\mu$. (Some of the excited states are nearly degenerate and their splitting is difficult to see in the figure.) The most important feature we find for the bare gold nanowire is the suppression of topological gap for higher chemical potential. It ranges from about $0.25\Delta$ to $0.08\Delta$ as $\mu$ changes from 12meV to 95 meV. At 500 meV, which corresponds to the chemical potential of surface states in gold, the gap is as small as $0.01\Delta$. Thus it is essential to reduce the chemical potential of the gold wire in order to detect the MZM's. In ref (1) it was proposed that this can be done via a gate on the gold surface. Fortunately, it turns out that EuS has the effect of reducing the chemical potential to about 15 mV. This makes the current experiment feasible.

The reason for the reduction in the energy gap for increasing chemical potential is mainly due to fact that the Rashba spin-orbital coupling $\alpha_R k_y \sigma_x$ dominates over the coupling $\alpha_R k_x \sigma_y$ . This last term normally pins the spin along the y direction for an electron moving along x and resists the Zeeman field's tendency to line up the spin parallel the magnetic field. Instead, we find that the Zeeman field easily forces the spins on opposite sides of the Fermi points of the last sub-band to be nearly parallel, and greatly reduce the coupling to the spin singlet pairing induced by vanadium. To see this in more details, we consider the realistic case where the width of gold nanowire is about to 50~100 nm, which is comparable to $l_{so}$. In this case, the electrons can have a transverse momentum before scattering by the boundary. The topological gap $E_G$ is controlled by the last sub-band which has the smallest longitudinal momentum $k_x$, but largest transverse momentum $k_y$ at the Fermi energy. For large chemical potential, $k_y$ is very large. The Rashba spin-orbital coupling $\alpha_R k_y \sigma_x$ is large and it has the effect of pinning the spins along the x direction so that a Zeeman field will line up the spins at the Fermi level to be nearly parallel to the magnetic field. In this case, the gap opened by the singlet s-wave pairing will be small. Evidently, this transverse Rashba SOC pinning effect is stronger when the chemical potential is increased, as we have observed in the simulation.

# 4 MZM in the partially covered island

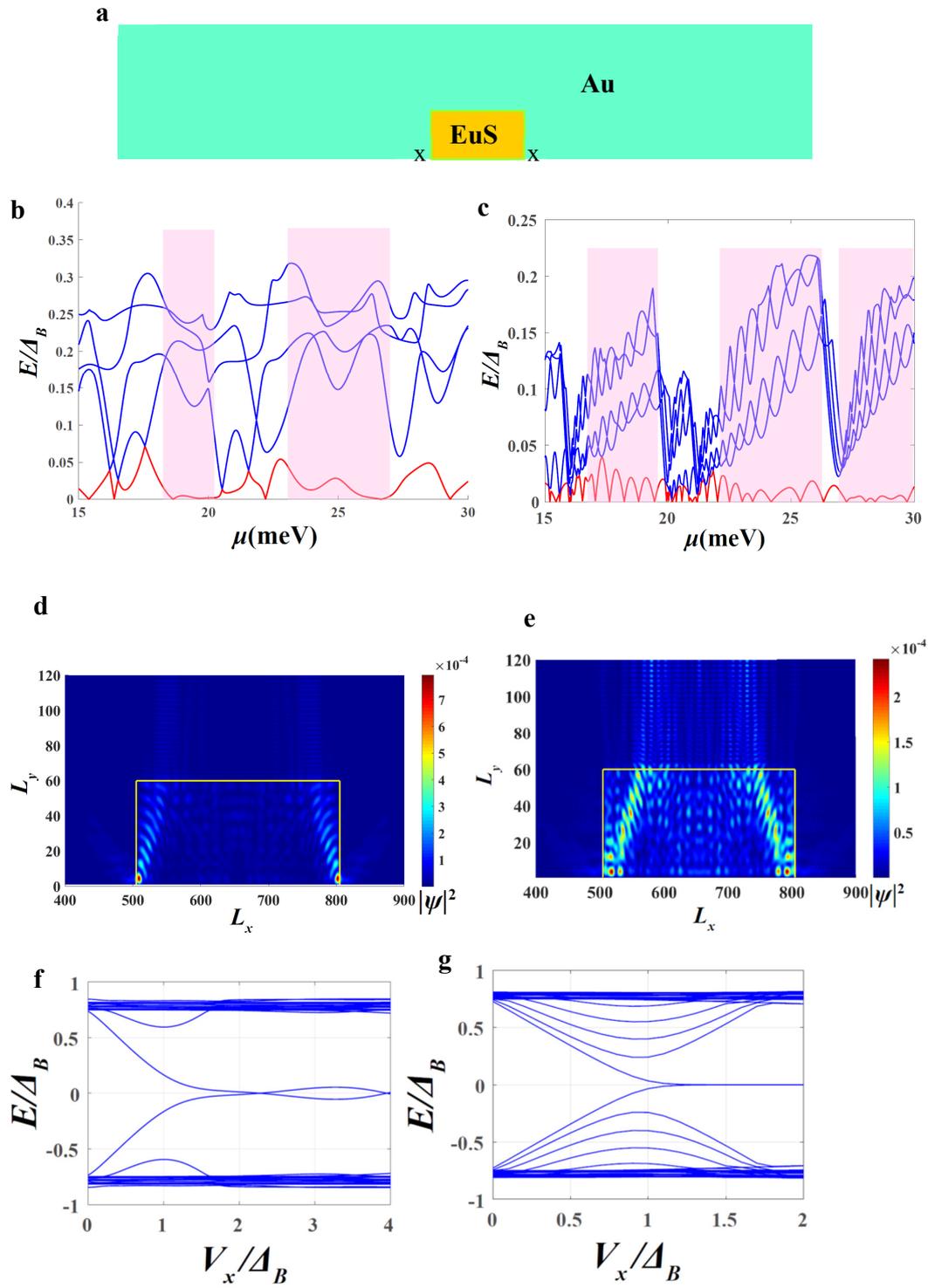

**Figure S2 | MZM pairs in the gold surface partially covered by EuS island.** The results are calculated from the Rashba wire model including the self-energy term with $\Gamma = 3\Delta_B$. The bulk superconducting gap $\Delta_B = 0.5$ meV and $\mu$ is chosen to be 25 meV under the island. **(a)** Schematic plot of a gold wire partially covered by an EuS island considered in the simulation. In the simulation, $\mu$ is set to be 200 meV, $V_x=0.2\ \Delta_B$ under bare Au region, and the effective Zeeman energy including exchange with the EuS is set to be $V_x=2\ \Delta_B$ under EuS. The two X points label where the ZBPs appear in the experiment. **(b)**, **(c)** show the lowest few eigenstates at the X points for a 60 nm x 60 nm and a 60 nm x 300 nm island respectively. The topological regions are shown by pink shading. **(d)** and **(e)** show the absolute value of the wavefunction for the MZM and an excited state at energy 0.251 $\Delta_B$ respectively for the 300 nm long island. (e) and (f) show the lowest few eigenvalues of a narrow gold wire (10 nm wide) with a 5nm wide EuS island for a 100nm and 300 nm long sample respectively vs $V_x$. $\mu$ is taken to be 45 mV and a single transverse mode is supported. Compared with the wider wire shown in Fig 5a there are very few low lying states and for the short wire, a single state can be seen closing and re-opening the gap, consistent with ref (10) in the main text,

This section discusses the results of our simulation when EuS island is placed on top of Au wires. We employ the method and model explained in sections S1 and S2. As discussed in section S3, we need to reduce $\mu$ in order to have a sizable topological gap. In the experiment, this can be done by covering the gold surface with the EuS islands as shown in **(a)**. Within the EuS covered region, $\mu$ can be lower than 30 meV as found in the experiment. In order to describe this new geometry, we consider the following improvements on the Rashba wire model used in the previous section. It turns out that three new ingredients are needed to produce MZM's confined to the islands. First, $\mu$ should be spatially dependent, which is set to be much lower under the EuS island than in the gold wire. In the simulations we choose the chemical potential in the gold region to be 200 mV with a transition region of width 5 nm that interpolate linearly to the μ value under the EuS. Second, we include the coupling of the surface state to the bulk gold states which we treat as a self-energy term. Thus the conductance calculations are based on Eqs. S5 and S6 with the full frequency dependence Z factor. The wavefunctions, on the other hand, are obtained from the effective Hamiltonian S16 which treats the Z factor as a constant. As discussed in the text, the Z factor helps us localize the MZM. The coupling strength $\Gamma$ plays an important role. In the calculation, we choose $\Gamma = 3\Delta_B$. Since the surface pairing gap seen in the experiment approaches the bulk gap $\Delta_B$, $\Gamma$ cannot be too small. The $\Gamma$ chosen gives a surface pairing gap consistent with what is seen in the experiments. On the other hand, a much larger $\Gamma$ is also not acceptable because Z is small and the low-lying excitation energies are renormalized by the factor Z. As such, we will have many low energy excitations very close to the MZM energy.

Third, we need to enhance the Zeeman energy under the island. Physically, this is due to exchange interaction with the ferromagnetically ordered magnetic moment in EuS. This turns out to be essential in order to trap the MZM in the island. This local enhancement of Zeeman energy not only enables the islands to be driven into the topological region but also keeps a sizable superconducting gap in the bare gold region, which confines the MZM's at the boundary of the islands. If this is not the case, a Zeeman energy large enough to drive part of the sample under EuS to be topological (for example, $V_x > \Delta_s$ where $\Delta_s$ is the induced gap on the surface) will also greatly reduce that pairing gap of the gold wire. Moreover, if the gold wire is extended to become two-dimensional, the pairing gap of the gold wire will be closed and the MZM's will not be found. In the experiment, 5T field is not sufficient to overcome $\Delta_s$ to drive a topological transition, so we must rely on the exchange enhancement of $V_x$ under EuS to drive that part of the sample topological.

The results of this model are shown in Fig. S2. We consider the same geometry used in the experiment, where the EuS island only overlaps with one side of gold wire as shown in **(a)**. We assume $V_x$ =0.2 $\Delta_B$ in the bare gold and 2 $\Delta_B$ under the EuS. Indeed, we find that the MZM's can be trapped at the islands after driving the islands into the topological regime. In a finite length wire the topological region is not a sharply defined concept, but we use the term to denote regions where the bulk gap has closed and a pair of split MZM appears. These are shown by pink shading in Fig. S2b and S2c for the 60nm x 60 nm and 60nm x 300 nm islands respectively. We see that compared with the gold wire without EuS shown in Fig S1, the probability of finding the topological region is quite high. Thus we expect to find MZM in the experiment a reasonable fraction of the time. We find distinct ZBPs near X points, which originate from the wavefunction of MZM being concentrated at the "hot spots" as indicated in **(d).** Note that the wavefunction leaks out of the hot spot mainly along the edge of the gold wire. This feature matches the experimental finding Shown in Fig 3, where ZBP are seen only at the corners where the EuS island meets the edge of the gold wire.

Note that many in-gap states are generated in the topological regime, as seen from the eigenvalue plots in Fig S2b and c. Typically these are separated from the MZM's by about 0.1 $\Delta_B$. Upon thermal averaging, these give rise to a conductance background which fills in the gap and also a shoulder to the ZBP. These features are consistent with the experimental data shown in the main text. Upon a more detailed examination of the eigenstates, we find that the lowest excited state and many of the higher energy ones are extended throughout the EuS island, but a number of the excited states are localized near the ends of the islands. **(e)** shows the wavefunction of a representative state at $V_x$ =2 $\Delta_B$ at energy 0.251 $\Delta_B$. It is interesting that the wavefunction strongly resembles that of the MZM shown in (d). We interpret this as a "Majorana descendant". As we can see from (b) and (c), every time we enter a non-topological region when the chemical potential is increased, the MZM moves up in energy to form a conventional in gap quasi-particle state. As a result, some of the low lying in gap state retains memory of the MZM wavefunction that they descended from. These states contribute to the tunneling spectra near the ends of the island but not in the middle.

The Majorana descendants and the large number of in gap states are properties of the multi-mode problem. To show the contrast, we show in (f) and (g) the lowest states for a very narrow wire that is in the single mode regime. For a short wire (f) we can see a single state descending in energy with increasing Zeeman field to close the gap and re-opens in the topological region. For longer wires we find a group of states. This is expected because in the infinitely long wire limit we expect a band of in gap state with momentum along the wire as a good quantum number. The single state shown in (f) is a result of length quantization. The gap closing by a single state appears both in the simulation and the experimental data in the recent paper on semiconductor nanowire and corresponds to the short wire case. (ref 10 in the main text.)

## 5. MZM in a large area of gold.

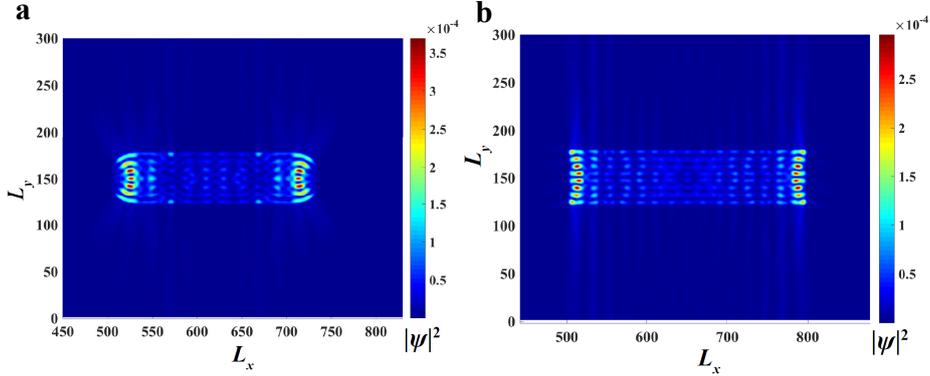

**Figure S3 | MZM's in a wide gold wire.** In **(a)** round corner island and **(b)** rectangular island, the localized MZM still exists in the 300 nm wide gold wire**.** Since only 3 to 4 T is needed in the experiment to drive the EuS covered island into the topological region, the Zeeman energy in the bare gold is smaller than $\Delta_B$, roughly only about $0.2\Delta_B \sim 0.3\ \Delta_B$, using a g factor of 2. However, the effective Zeeman energy is large under the island due to exchange interaction with the moments in EuS. In this case, we find MZM's at the end of the island even for a planar gold surface. Therefore, we believe it is feasible to simply evaporate the EuS on a large area of gold and do away with the gold wire completely.

Finally, we briefly discuss placing an island in the middle of a relatively wide gold wire, in order to simulate the situation of creating EuS island on a gold surface, without having the need to make a narrow gold wire. As shown in Fig. S3 we find that MZM's are indeed created at the ends of the EuS islands, provided that the Zeeman energy is enhanced under EuS. However, there are two caveats. The MZM's are spread out more or less uniformly across the ends of the islands, thus reducing the coupling with the STM tip. As a result, the size of the ZBP is small relative to the background. This issue may be overcome by replacing the STM with a planar junction that covers the ends of the EuS island. Second, for longer islands, we find an energy gap of about $0.1\ \Delta_B$ in the tunneling spectrum, but we find many states above the energy gap. Thus, in order to observe the ZBP due to MZM's, a temperature resolution better than $0.1\ \Delta_B$ will be needed.

## 6 Derivation of self-energy term from surface-bulk mixing

For simplicity we assume that the gold surface states and bulk states are mixed by impurity scattering.[1] The parameter $\Gamma$ can parametrize other forms of mixing such as virtual exchange of phonons or screened Coulomb intereaticon. The action can be written as

$$S[\bar{\psi}_B, \psi_B, \bar{\psi}_S, \psi_S] = \int d\tau \int d^3 r \bar{\psi}_B(\mathbf{r},\tau)(\partial_\tau + \widehat{H}_B + V(\mathbf{r})\tau_z)\psi_B(\mathbf{r},\tau) +$$

$$\int d\tau \int d^2 \mathbf{r}_\parallel \bar{\psi}_S(\mathbf{r}_\parallel,\tau)(\partial_\tau + \widehat{H}_S + V(\mathbf{r}_\parallel, z=0)\tau_z)\psi_S(\mathbf{r}_\parallel,\tau) + \int d\tau \int d^3 \mathbf{r}$$

$$\bar{\psi}_B(\mathbf{r},\tau)V(\mathbf{r})\delta(z)\tau_z\psi_S(\mathbf{r},\tau) + \bar{\psi}_S(\mathbf{r},\tau)V(\mathbf{r})\delta(z)\tau_z\psi_B(\mathbf{r},\tau), \tag{S9}$$

where $\psi_{B(S)}$ is the bulk (surface) Fermion field, $V(r)$ is the disorder potential. The disorder is random, $\overline{V(r)} = 0$, and has short correlation $\overline{V(r)V(r')} = W^2\delta(r - r')$. The disorder average is calculated with a Gaussion weight, $\overline{(\cdots)} = \int D[V](\cdots)\exp(-\int dr \frac{V^2(r)}{2W^2})$. After integrating out the bulk field, we can obtain an effective action to describe the surface states. To do this, we formulate the action in the frequency $\omega_n$ and momentum $\mathbf{k} = (\mathbf{q}, k_z)$ space, and integrate out the bulk Grassmann fields using

$$\int d(\bar{\phi}, \phi) e^{-\bar{\phi}^T A \phi + \bar{v}^T \cdot \phi + \bar{\phi}^T \cdot v} = \det A e^{\bar{v}^T \cdot A^{-1} \cdot v}. \tag{S10}$$

We can see that the effective action becomes:

$$S_{eff}[\bar{\psi}_S, \psi_S] = \sum_n \sum_{q,q'} \bar{\psi}_S(\mathbf{q}, \omega_n)(-i\omega_n + \hat{H}_s(\mathbf{q'})\delta_{\mathbf{q},\mathbf{q'}} + V(\mathbf{q} - \mathbf{q'}, z = 0)\tau_z)\psi_S(\mathbf{q'}, \omega_n)$$

$$+ \sum_n \sum_{q',q''} \bar{\psi}_S(\mathbf{q'}, \omega_n) \sum_k V(\mathbf{q'} - \mathbf{q}, z = 0)\tau_z G_B(\mathbf{k}, \omega_n) V(\mathbf{q} - \mathbf{q''}, z = 0)\tau_z \psi_S(\mathbf{q''}, \omega_n).$$
(S11)

Then, we average over all the disorder configurations within the Born approximation:

$$\overline{S_{eff}}[\bar{\psi}_S, \psi_S] \approx \sum_n \sum_{q,q'} \bar{\psi}_S(\mathbf{q}, \omega_n)(-i\omega_n + \hat{H}_s(\mathbf{q'})\delta_{\mathbf{q},\mathbf{q'}} + V(\mathbf{q} - \mathbf{q'}, z = 0)\tau_z)\psi_S(\mathbf{q'}, \omega_n)$$

$$+ \sum_n \sum_{q'} \bar{\psi}_S(\mathbf{q'}, \omega_n) W^2 \sum_k \tau_z G_B^{(0)}(\mathbf{k}, \omega_n) \tau_z \psi_S(\mathbf{q'}, \omega_n), \tag{S12}$$

where $G_B^{(0)}(\mathbf{k}, \omega_n) = (i\omega_n - \hat{H}_B(\mathbf{k}))^{-1} = (i\omega_n \tau_0 - \xi \tau_z - \Delta_B \tau_x - \mathbf{V} \cdot \boldsymbol{\sigma} \tau_0)^{-1}$. So the self-energy term is

$$\Sigma(i\omega_n) = W^2 \sum_k \tau_z G_B^{(0)}(\mathbf{k}, \omega_n)\tau_z \approx -\Gamma \frac{(i\omega_n - \mathbf{V} \cdot \boldsymbol{\sigma})\tau_0 - \Delta_B \tau_x}{\sqrt{\Delta_B^2 + \omega_n^2}}, \tag{S13}$$

where $\Gamma = \pi N_B(0) W^2$, $N_B(0)$ is the density states at the bulk Fermi surface. In the last step, we neglect the Zeeman dependent term in the denominator, considering the Zeeman energy to be much smaller than the bulk pairing gap, otherwise superconductivity will be killed in the bulk. As discussed later, this is certainly justified when the effective Zeeman energy under the EuS region is enchanced due to exchange intereaction with the Eu moments. After incoporating the self-energy term, the surface-state Green's function is

$$G(\mathbf{k}, i\omega_n) = 1/\{i\omega_n \tau_0 - (\frac{k^2}{2m} - \mu)\tau_z - \alpha_R(\mathbf{k} \times \boldsymbol{\sigma}) \cdot \hat{z}\, \tau_z - \mathbf{V} \cdot \boldsymbol{\sigma} \tau_0$$
$$+ \Gamma \frac{(i\omega_n - \mathbf{V} \cdot \boldsymbol{\sigma})\tau_0 - \Delta_B \tau_x}{\sqrt{\Delta_B^2 + \omega_n^2}}\}$$

$$= Z/\{(i\omega_n - \boldsymbol{V}\cdot\boldsymbol{\sigma})\tau_0 - Z\left(\frac{k^2}{2m}-\mu\right)\tau_z - Z\alpha_R(\boldsymbol{k}\times\boldsymbol{\sigma})\cdot\hat{z}\,\tau_z - (1-Z)\Delta_B\tau_x\}, \tag{S14}$$

where the quasiparticle weight is

$$Z(i\omega_n) = (1 + \frac{\Gamma}{\sqrt{\Delta_B^2+\omega_n^2}})^{-1}. \tag{S15}$$

In the low energy limit, we can drop $\omega_n$ in Eq (S15) and obtain the effective Hamiltonian

$$H_{eff} = Z(\frac{k^2}{2m}-\mu)\tau_z + Z\alpha_R(\boldsymbol{k}\times\boldsymbol{\sigma})\cdot\hat{z}\,\tau_z + (1-Z)\Delta_B\tau_x + \boldsymbol{V}\cdot\boldsymbol{\sigma}\,\tau_0, \tag{S16}$$

where

$$Z \approx (1+\frac{\Gamma}{\Delta_B})^{-1}. \tag{S17}$$

In our simulation we use the full frequency dependence of the self-energy as given in Eqs. S5 and S6, except when we calculate the wavefunction of the low energy eigenstates as shown in Fig 4 where H$_{eff}$ is used. It is worth noting that the kinetic energy (Fermi velocity) and the Rashba energy are renormalized by $Z$, while the Zeeman energy is not. Physically, it is because the surface states and bulk states experience the same Zeeman field. On the other hand, it is important to note that Eq. S16 applies only to energy levels much less than the bulk gap $\Delta_B$. As a consequence, when $\Gamma$ is much larger than $\Delta_B$, which is the case of interest, the actual surface energy gap $\Delta_S$ is close to $\Delta_B$ and Eq. (S16) is no longer applicable. $\Delta_S$ is not simply given by $(1-Z)\Delta_B$. In order to determine the surface gap $\Delta_S$, it is necessary to keep the $\omega_n$ dependence of $Z$ in Eq. S15 and solve for the smallest pole in Eq. S14.[1] In the strong coupling limit, $\Gamma \gg \Delta_B$, the asymptotic behavior of the surface gap is $(1-\frac{\Delta_B^2}{\Gamma^2})\Delta_B$, which is comparable to the bulk pairing gap in our situation. In practice we find that with the parameters used in this paper, the effective Hamiltonian is accurate for energy less than 0.3 $\Delta_B$.

Next, we consider the effective Hamiltonion of the surface state under the EuS island. The main effects of EuS are locally shifting $\mu$ and enhancing the Zeeman energy under the EuS island. The latter effect comes from the exchange interaction with the ferromagnetically ordered magnetic moments in EuS. The total Zeeman energy under the islands can be written as $V_{ex}Z_i + V_{0x}$, where $V_{ex}$ is the Zeeman energy induced on the surface states due to EuS, and $V_{0x}$ is the Zeeman energy due to the applied in-plane magnetic field in the $x$ direction. Here $V_{ex}$ is renormalized by a factor $Z_i$ because the electrons in the surface state under EuS spend a fraction of the time in the bulk gold. In the calculation in order to avoid having too many parameter, we replace $Z_i$ by its zero frequency limit and we write $V_{ex}Z_i + V_{0x}$ as an effective enhanced Zeeman energy $V_x$. This is used as a parameter throughout this paper. For example, in section S4 and S5, we set this energy as $2\Delta_B$.

# 7 Lithographically patterned Au(111) nanowire array

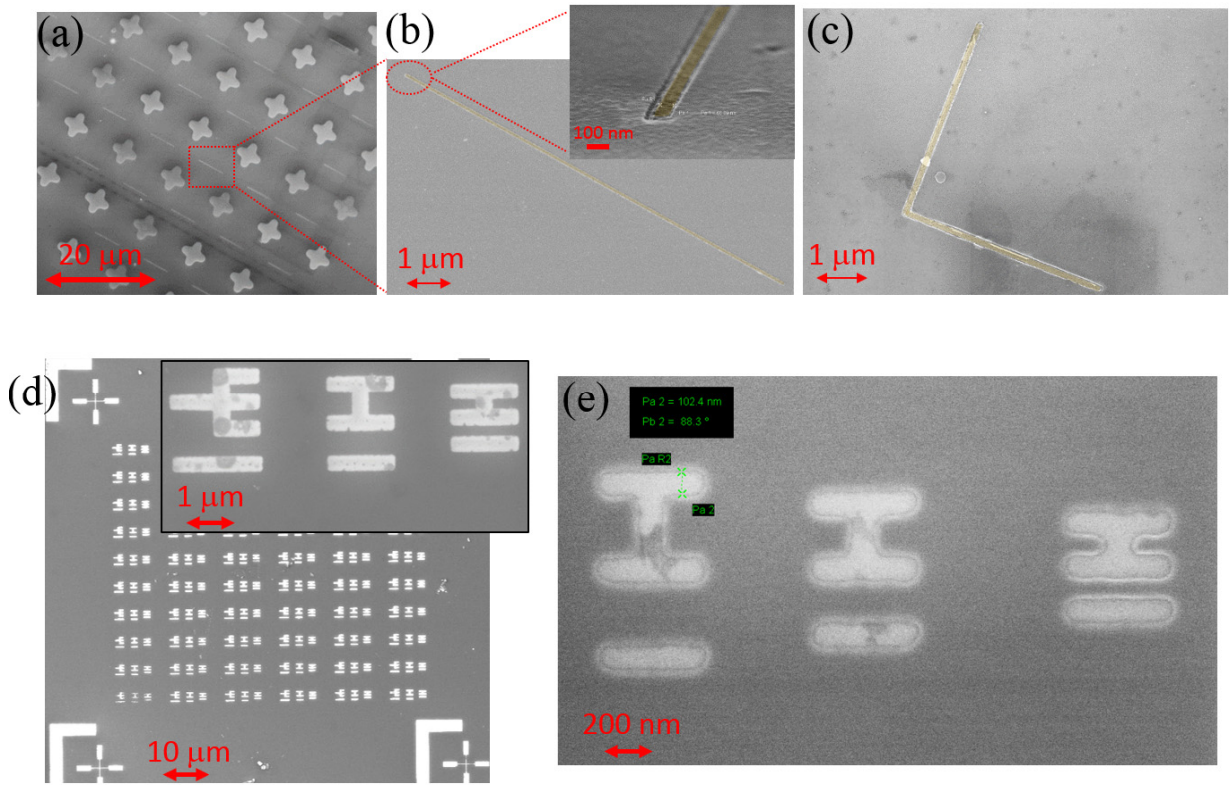

**Figure S4 | Scalable nanowire array (a)** The scanning electron microscopy (SEM) of an array of Au(111) nanowires. Each nanowire is sitting in a square area with four alignment marks (cross) at the corners. **(b)** The zoomed-in SEM image of the region marked by the red square in **(a)**. The inset shows the zoomed-in SEM image of the area marked by an oval. The SEM is taken is a tilted way so that the vertical layout of the nanowire can be seen. **(c)** A "L" shape nanowire fabricated using similar method. **(d)** A patterned array of the "tetron" structure for MZM qubits. Inset shows a typical unit of such array. **(e)** The patterned "H" bar structures for MZM qubits. The dark patch regions in **(d)** and **(e)** are solvent residues left during the nanofabrication, which can be easily removed by following standard cleaning procedures.

Scalable Au(111) nanowire array is fabricated out of the epitaxially grown Au(111) thin films (Fig S4 (a)). The standard lithography is combined with the custom designed nanofabrication processes to achieve such nanowire array. Au(111) nanowires with width ~ 70 nm and length ~ tens of μm is made (Fig S4 (b)). The Au(111) layer is sitting on vanadium film and fabricating the Au(111) nanowires inevitably exposes the vanadium surface, which is known to be vulnerable to oxidation. Such oxidation will be detrimental to the induced superconductivity into the Au(111) nanowire. Therefore, special procedures are taken to protect the exposed vanadium surface during the nanofabrication, which has been proven successful as can be seen by the large induced superconducting gap (comparable to that of bulk vanadium) in the Au(111) nanowire (Fig 1d main text). Our nanofabrication also maintains the atomic clean Au(111) surface even after being through all the lithography procedures. Special shapes of Au(111) nanowire (Fig S4 (c) "L" shape nanowire) can be made. In addition, H shaped and more complex structures called "tetrons" that can trap four MZM at the 4 ends (proposed in ref. 34 in the main text as building blocks for qubits) have been made in order to demonstrate the flexible design using the software for standard lithography. (see Fig. 5S (d), (e)) Such flexibility allows us to build more complex Au(111) nanowire circuits for manipulating MZM's.

**8 The Spatial decay of MZM.**

The spatial decay of MZM is measured near position 1 (Fig S5a) as a function of the distance as the tip is moved to locations away from the island. The direction of the scan is marked by the small arrow next to the label of position 1 (Fig S5a). The ZBP loses about half its height when the tunneling location is ~ 8 nm away from the edge of the EuS island and the full gap is restore at about 14 nm. We thus define the decay length $\xi \sim 8$ nm.

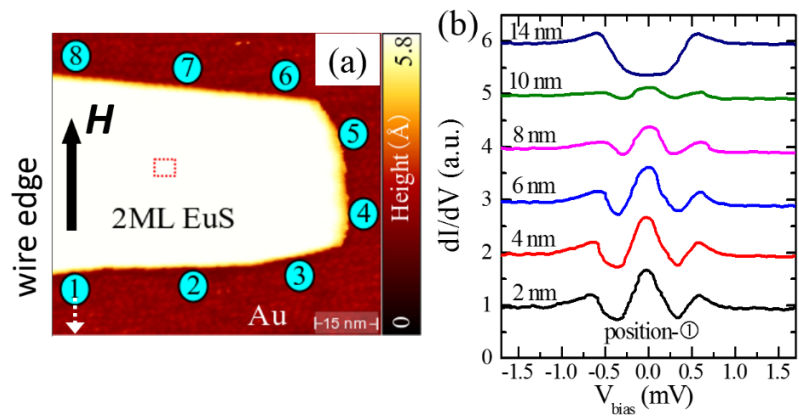

**Figure S5 | Spatial decay of MZM** (a) The same island as shown in the main text Fig 4b. The white dotted arrow next to position 1 indicates a typical direction of various dI/dV scans for the study of the decay of ZBPs. Note that here the demonstrated scans (STS shown in (b)) are along the Au(111) nanowire edge. It thus rules out the possibility that the ZBPs were due to impurities located at the edge of the Au(111) wire. (b) A series of dI/dV spectra taken at different locations along the white arrow noted next to position 1 in (a). The bottom most curve is taken at position 1, which is ~ 2nm away from the edge of the EuS island. The length noted on each spectrum indicates the distance between the measurement location and the edge of the EuS island. Each curve is normalized to 1 at the highest bias voltage and is shifted by one unit along the y-axis for clarity.

**9 The superconducting gap of vanadium film itself as a function of the magnetic field**

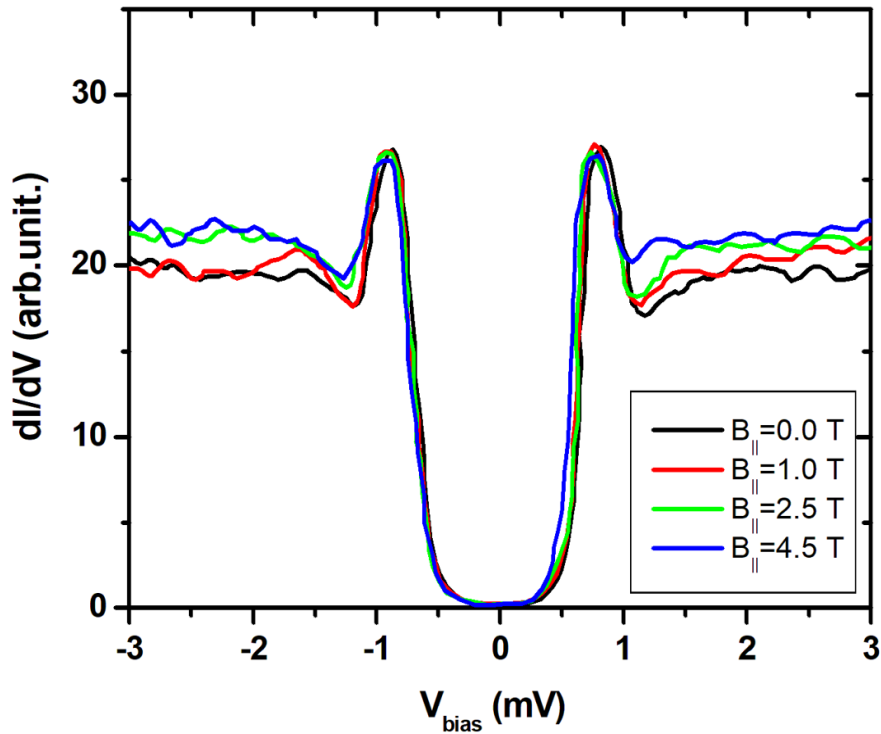

**Figure S6 | The SC gap of vanadium as a function of the applied magnetic field.** The vanadium has a hard superconducting gap. The applied magnetic field does not create any in gap states. We also note that the tunneling conductance is zero at zero bias voltage.